# Quantum-chemical perspective of nanoscale Raman spectroscopy with three-dimensional phonon confinement model

Vitaly I. Korepanov[1,2], Hiro-o Hamaguchi[1]*

1. National Chiao Tung University, 1001 University Road, Hsinchu, Taiwan 30010
2. Institute of Microelectronics Technology and High Purity Materials, RAS, Chernogolovka 142432, Russia

e-mail: hhama@nctu.edu.tw

**Raman spectroscopy of crystalline/molecular systems is well backed with quantum chemical calculations and group theory, making it a unique characterization tool. For the "intermediate" case of nanoscale systems, however, the use of Raman spectroscopy is limited by the lack of such theoretical bases. Here, we suggest to couple a scaled quantum-mechanical (SQM) calculation with the phonon confinement model (PCM) to construct a universal and physically consistent basis for nanoscale Raman spectroscopy. Unlike the commonly used one-dimensional dispersion PCM, we take into account the confinement along all the three dimensions of the $k$-space. We apply it to diamond nanoparticles of sub-50nm size, a system with pronounced anisotropy of dispersion for which the use of three-dimensional dispersion is a requisite. The model excellently reproduces size-sensitive spectral features, including the peak position, bandwidth and asymmetry of the $sp^3$ C-C Raman band. This fundamental approach can be easily generalized to other nanocrystalline solids to hopefully contribute the future development of quantitative nanoscale Raman spectroscopy.**

The continuously growing interest to nanoscale matter drives a strong demand for fast, non-invasive and statistically reliable characterization methods. This makes Raman spectroscopy an indispensable tool for nanoscience.

The difficult challenge of Raman spectroscopy of nanoscale materials is to establish a consistent way to link the Raman pattern to the characteristic size of the crystallites.

The straightforward approach is to directly calculate the normal modes of nanoparticles as big molecules[1,2]. This requires assumptions of the shape, size and symmetry. The calculated spectra are the set of narrow lines. Since the number of normal modes for nanoparticles is in the order of thousands, such calculations do not seem to be routinely availbale for real systems.

The other common approach is the elastic sphere model (ESM), in which the nanoparticle is approximated by a homogeneous isotropic free-standing elastic sphere. The vibrations of such system can be easily derived from first principles calculations[3]. For the nanoparticles of other shapes, as well as systems with the size smaller than ~5nm, the applicability of ESM is limited. For more details on ESM the reader is referred to the review[4].

The third general approach starts from the Raman spectrum of a bulk crystal, for which the selection rules allow only the phonons close to the Brillouin zone (BZ) center. Nano-size crystallites lack the translational symmetry and the quasimomentum preservation does not hold. As a result, BZ points away from the center can also contribute to the Raman spectra. To what extent the selection rule breaks depends on the size of the crystallite. This effect is named the "phonon confinement" after the work of Richter[5], who first suggested the physical model for describing it.

The approach by Richter's is to multiply the phonon wavefunction of the crystal $\varphi(q_0,r)=u(q_0,r)*\exp(i\mathbf{q_0}\mathbf{r})$ with the confinement function $W(r,L)$ that reflects the crystallite size ($L$).

The resultant wavefunction is not any more an eigenfunction of the phonon wave vector $q$, but rather is a superposition of crystal wavefunctions with different $q$. This superposition is a reflection of the "mixing" of the phonon modes of the crystal upon confinement. Therefore, the "smaller" the crystallite size is, the more is the "spreading" of the phonon wavefunction over the BZ. The weights of different $q$-points within this superposition are determined as the squared Fourier coefficients $C(q_0,q)$ of the confined wavefunction expansion:

$$\psi(q_0,r) = u(q_0,r) * \int C(q_0,r) * e^{iqr} d^3q \quad . \quad (1)$$

$C(q_0,q)$ is specific to the choice of the confinement function $W(r,L)$:

$$C(q_0,q) = \frac{1}{(2\pi)^3} \int W(r,L) * e^{-i(q-q_0)r} d^3r \quad . \quad (2)$$

The Raman band shape is given as a superposition of the contributions from different $q$, weighted by the Fourier coefficients. If the Lorentzian line shape is assumed for each $q$-point, the following expression for the band shape $I(\omega)$ is obtained:

$$I(\omega) \cong \iiint \frac{\left|C(q_0,q)^2\right|^2}{(\omega-\omega(q))^2 + (\Gamma_0/2)^2} d^3q \quad , \quad (3)$$

where $\omega$ is wavenumber, $\omega(q)$ is the three-dimensional (3D) phonon dispersion function, $\Gamma_0$ is the natural bandwidth. This is the general formula for the Raman band shape under the phonon confinement model.

The PCM received criticism for assuming the isotropy of phonon dispersion, as well as the arbitrariness of the confinement function $W(r,L)$, because of which the PCM was considered as "phenomenological"[6]. Also, it was believed to be not applicable to particles below ~5 nm, where the bulk dispersion curves lose their significance[4].

In the present manuscript, we show the way to exclude the arbitrariness in the confinement function, and take into account the full anisotropy of the phonon dispersion.

As of the size limit, the main advantage of PCM is that it has perfectly good asymptotic behavior on "both ends" of the size scale. For large crystallites the calculated spectrum coincides with that of the bulk crystal, and for the case of a disordered (amorphous) material the spectrum uniformly approaches the density of states (which corresponds to equal contributions from all points of the BZ or, in other words, constant Fourier coefficients in equation 3). The experimental spectra of amorphous solids (system with the smallest phonon coherence length in terms of PCM) are known to closely resemble the density of states[7].

In the past, in order to simplify PCM, the 1D approximation has often been introduced. under which the 3D dispersion function was replaced with an averaged 1D dispersion curve. However, in case of distinct anisotropy of phonon dispersion like over-bending[8], one-dimensional approximation curve leads to very arbitrary results. A typical example is diamond, for which several groups suggested different 1D dispersion curves, neither of which can really approximate the 3D function[9]. Therefore, the consistent physical model should adopt the 3D integration all over the $q$-space, which requires the knowledge of the phonon dispersion at any arbitrary $q$-point. Here, we greatly benefit from quantum-chemical calculations.

Typically, quantum-chemical calculations tend to under- or over-estimate experimental vibrational frequencies more or less systematically[10]. To take this into account, we use the scaling procedure, routinely used for molecular systems. In our case, we scale the 3D dispersion function to match the experimental frequency at the BZ center. If the phonon dispersion is degenerate, like that of diamond or silicon, we need only one scaling factor.

We illustrate the present approach by applying it to diamond nanoparticles. They comprise an excellent model system due to the high phonon dispersion, simplicity of the first order spectrum, and the availability of samples with different crystallite size. In addition, the phonon dispersion of diamond is well studied, which allows us to compare the calculated frequencies with high-quality experimental data[8,11–14].

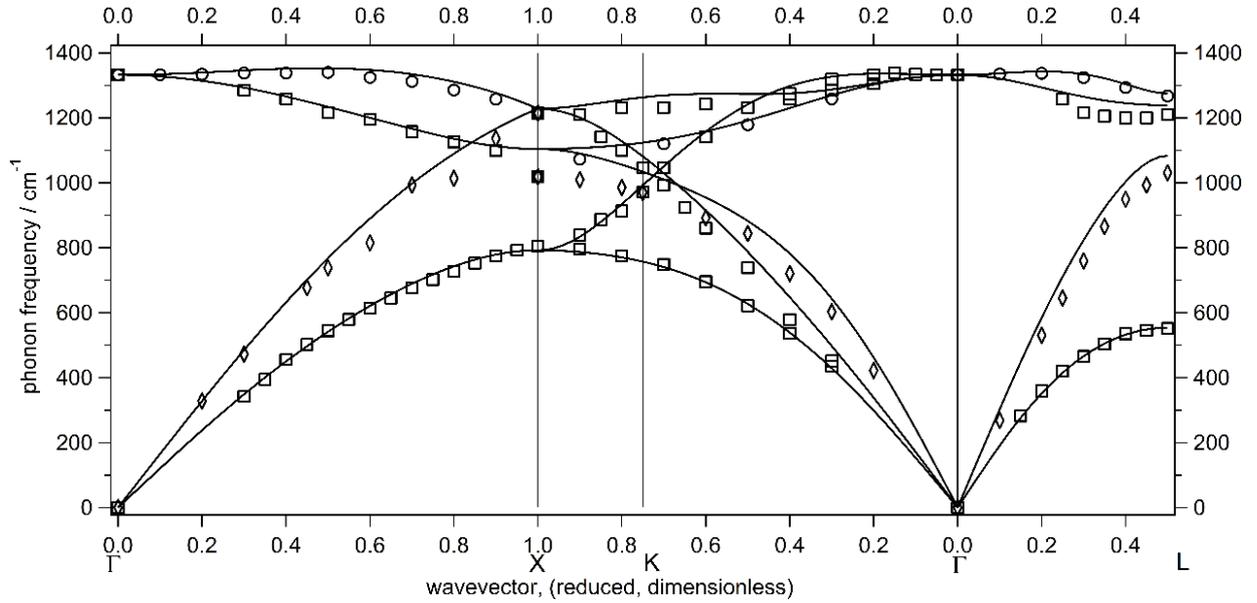

**FIG. 1**. Calculated and experimental phonon dispersion functions of diamond plotted along high-symmetry directions. Lines: scaled quantum-mechanical calculation results, circles: X-ray scattering data[8,13]; squares: neutron scattering data[11,12]; diamonds: neutron scattering data[14].

The SQM calculated phonon dispersion function of diamond is plotted along high-symmetry directions in fig. 1. X-ray scattering data[8,13] and neutron scattering data[11,12,14] are also shown in the same figure. The original DFT calculation was found to under-estimate the frequency at BZ center point by 27 cm$^{-1}$, which corresponds to 1.02 scaling factor. After the scaling, the agreement between calculated and experimental data is good throughout the BZ. Therefore, we assume that any arbitrary $q$-points are also well reproduced by the SQM calculation. We then directly use this scaled 3D dispersion function $\omega(q)$, defined on a grid, for the integration over $d^3q$.

It is worth mentioning that the phonon confinement model has been considered as phenomenological for a long time, because the phonon weighting function has been chosen somewhat arbitrarily, without physical justifications[6]. Typically, a Gaussian has been assumed for the simplicity of the integration[9]. In the present work, however, no assumptions were made for the weighting function. The Fourier coefficients were derived directly from the confinement shape. Since the diamond nanocrystals can be well described as quasi-spherical[15], we use a spherical box confinement, and $C(q_0,q)$ take the analytic form of Bessel-like function[16]. The present approach is applicable to any arbitrary confinement shape. If the analytic expression for $C(q_0,q)$ does not exist, it can be calculated numerically[17]. As illustrated in fig.2, the exact phonon weighting functions are significantly different from those derived assuming the Gaussian confinement.

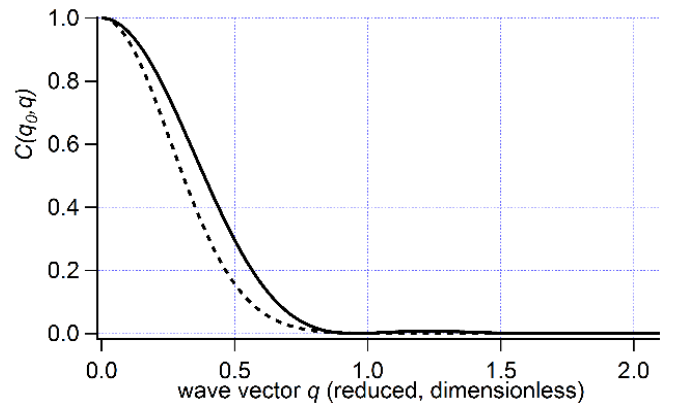

**FIG. 2.** Fourier coefficients for the hypothetical small nanoparticle (3 lattice units) derived from the Gaussian confinement function (dashed) and the exact one (solid).

The Raman cross-section is proportional to the cube of the particle size; therefore, the contribution of the different-size particles is not uniform. For this reason, it is necessary to take into account the

particle size distribution. This makes the working formula:

$$I(\omega) \cong \int \rho(\sigma)d\sigma \frac{\sigma^3}{N(\sigma)} \iiint \frac{\Gamma_0(\sigma) * |C(q_0,q)^2|^2}{(\omega - \omega(q))^2 + (\Gamma_0(\sigma)/2)^2} d^3q \quad (4),$$

where $\rho(\sigma)$ is the particle size distribution and $N(\sigma)$ is the normalizing factor (the area under the spectral shape for the given size $\sigma$)[18]. The integration over $q$ is not limited to the BZ, because the lattice periodicity is broken[17]. In practice, it matters only for relatively small particles. Therefore, the integration can be safely limited to the finite area, in the present case extending the first BZ to 2.0 (fig. 2).

For comparison of the theoretical Raman spectra with the experimental, we need to include three more factors: (1) the frequency factor ($v^4$); (2) the Bose-Einstein factor; and (3) the slit-function of the spectrometer. For the third factor, the calculated Raman spectra $I(\omega)$ are convoluted by the spectrometer slit-function, for which the Gaussian function with FWHM of 7.0 cm$^{-1}$ was taken in the present work.

The experimental Raman spectra of the four diamond nanoparticles samples are shown in fig. 3. They exhibit sharp size dependence for the particle size smaller than 50 nm. The Raman spectrum of a bulk diamond shows a symmetric band with FWHM of 11.1 cm$^{-1}$. For the "small" 3nm particle (fig. 3 (a)), however, the band becomes asymmetrically broadened (58 cm$^{-1}$ FWHM), with a distinct down shift of peak position (4.5 cm$^{-1}$). The 42nm sample shows a spectrum very close to the bulk spectrum, with only difference in the band width (13.7 cm$^{-1}$, fig. 3 (d)). The other two samples with intermediate sizes show intermediate band shapes with FWHM of 15.8 and 46.4 cm$^{-1}$ (fig. 3 (c) and (b)). These size-dependent characteristics are well accounted for by the present model; all the four experimental Raman spectra (highlighted by thick lines on dotted lines) are excellently fitted by using eq. 4 (full lines in fig. 3).

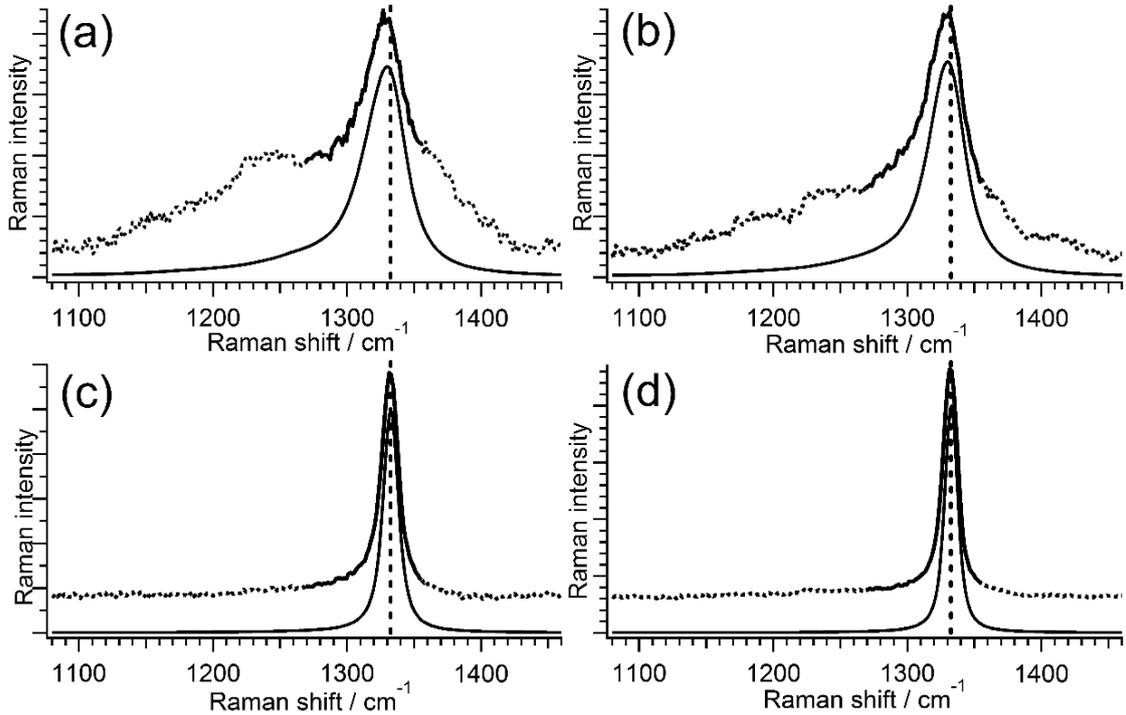

**FIG. 3**. Experimental Raman spectra (top dotted lines) and theoretical Raman spectra fitted by using eq.4 (bottom full lines) for 3nm DND (a), 4nm LND (b), 23nm HPHT (c) and 42 nm HPHT (d). Highlighted lines in the experimental spectra are the first order diamond band (1270-1355 cm$^{-1}$) that was used for fitting. Vertical dotted lines show the BZ center (1332.9 cm$^{-1}$).

The particle size distributions resulting from the fitting analysis are shown in fig. 4. Diamond nanoparticles typically consist of a diamond core surrounded by the shell of disordered sp3-, graphene-like carbon, fullerene-like islets as well as surface terminal carbon[19–21]. Hence, the phonon confinement length estimated in the present study is expected to be less than the particle size estimated from other methods. In fact, for sample (a), the present analysis gives a size distribution having a maximum at 2.8 nm (fig. 4 curve (a)), while dynamic light scattering (DLS) gives a size of 3.0 nm. Sample (b) shows the size distribution with the maximum of 3.4 nm (fig. 4 curve (b)), which is smaller than the value determined with XRD (4 nm). The bigger HPHT samples with 23.6 and 42.4 nm mean DLS sizes show good agreements with the mean confinement lengths of 22.2 and 42.0 nm (fig. 4, curves (c) and (d)).

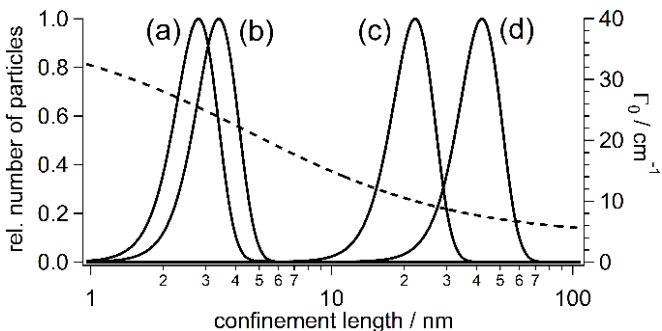

**FIG. 4**. Size distributions corresponding to the best fitted Raman spectra for samples (a), (b), (c) and (d) (solid lines) and the natural linewidth $\Gamma_0$ (broken line).

What PCM allows us to derive from the Raman pattern is the dimension of coherently scattering crystalline domain. It is worth emphasizing that this value does not necessarily coincide with the particle size. For the correct interpretation of the data, it is necessary to have the understanding of the structure of nanoparticles. In case of detonation nanodiamonds, TEM studies show that in the core, the nanoparticles have well-ordered diamond lattice. However, stacking faults are observed in a significant amount of nanoparticles[22]. Interestingly, the combined TEM+Raman studies have shown that the non-periodic arrangement of the stacking faults does not affect the phonon confinement length in SiC[23]. Therefore, it is likely that for the case of NDs the defects give minor contribution to the phonon coherence length, and the correct interpretation of the "Raman size" is the diameter of the diamond core.

Within the framework of the presented approach, the broadening of a Raman line of nanoparticles has two origins: (1) inhomogeneous broadening due to the mixing of the BZ-center phonon with off-center phonons having different frequencies, and (2) homogeneous broadening coming from vibrational dephasing. The former shows asymmetric broadening coming from the spreading of the phonon throughout the BZ with the generally negative phonon dispersion (fig.1). The latter would be dependent on the size of the confined crystallite. According to Nemanich and co-authors, the vibrational dephasing times are likely to be strongly influenced by surface and boundary scattering effects[24]. In the present work, we do not discuss the physical mechanisms involved in the homogeneous broadening as well as the analytical equation for it. We refrained from using the inverse size dependence of the dephasing constant $\Gamma_0$ as suggested in the work[25], because it gives unreasonably high $\Gamma_0$ values as the size approaches the amorphous solid. We optimized the size dependent $\Gamma_0$ values by fitting the theoretical band shape with the experimental spectra. In agreement with previous studies[24,25], $\Gamma_0$ has been found to increase with decreasing the size (fig.4, dotted line). This tendency is well expected from stronger perturbations that a phonon experiences in smaller particles.

Both quantum-mechanical calculations for periodic systems and PCM for nanocrystals are commonly used techniques. For diamond, the calculation of phonon dispersion can be done on a desktop computer within several days. For most other materials, this would require more time-demanding but still feasible calculations. Next step, running the PCM script takes several minutes. Therefore, once the phonon dispersion is calculated and scaled in the same manner, the approach can be easily applied to any other material systems.

In conclusion, coupling of quantum-chemistry with the phonon confinement model makes a universal and physically consistent basis for quantitative interpretation of the experimental Raman spectra from nanocrystalline solids. This method allows one to go beyond the 1D approximation to take into account the full anisotropy of the phonon dispersion. The present approach can easily be generalized to Raman spectral analysis of any other nanoscale materials.

**Materials and methods**

Raman spectra were recorded with a laboratory-built Raman system under 355 nm excitation. The de-focused laser beam (~1mm in diameter, ~60mW) was used; the signal was collected at 90° scattering geometry. The samples were in the form of the 1% diamond colloid in water. The rotating quartz cell was used in order to minimize the effects of laser heating.

The following four samples were measured:

**(a)** DND, detonation nanodiamond, with mean size of 3 nm (as estimated from DLS), courtesy of Prof. Eiji Osawa, NanoCarbon research Institute, Japan[26];

**(b)** LND, laser-synthesized ND with 4 nm mean size (as estimated by XRD), courtesy of Olga Levinson and Boris Zousman, Ray Techniques Ltd., Israel[27];

**(c)** HPHT, high-pressure high-temperature diamond, acid-washed and air-oxidized (3-4 hr at 450°C), with mean size 23 nm (as estimated by DLS),

**(d)** HPHT diamond, acid-washed and air-oxidized (3-4 hr at 450°C), with mean size 42 nm (as estimated by DLS), courtesy of Chandra Prakash Epperla and Prof. Huan-Cheng Chang, Institute of Atomic and Molecular Sciences, Academia Sinica, Taiwan[28].

All diamond powders were ultrasonicated in a horn for at least 0.5 hr, which yielded stable colloids (no precipitation within the observation time of about 1 year).

Quantum-mechanical phonon frequencies were calculated by means of Quantum Espresso software[29] with PBEsol-PAW functional and 16x16x16 $k$-points for SCF. The phonon frequencies were multiplied by one scaling factor (1.02) to match the experimental zone-center phonon frequency (1332.9 cm$^{-1}$).

For the calculation of the Fourier coefficients, the 3D step-function was used. Such approach has been used for a long time in crystallography for the description of nano-powder diffraction pattern[16]. It has been shown that for spherical nano-particles the Fourier coefficients take form of the Bessel function:

$$C(q_0,q) = \frac{4}{3}\pi R^3 \left[ 3\frac{\sin(qR) - qR*\cos(qR)}{(qR)^3} \right].$$

The fit of the experimental Raman spectra with the PCM (eq. 4) was made with the Matlab (2014b) script. Since the first order diamond line (around 1332.9 cm$^{-1}$) is superimposed by the disordered sp3 carbon line (broad feature at 1240 cm$^{-1}$) and D-band of graphene-like carbon (1368 cm$^{-1}$), the fit was made only for the 1270-1355 cm$^{-1}$ region, where the first order diamond line is the most distinct. The fitting variables included the parameters of the size distribution, as well as the natural linewidth $\Gamma_0(\sigma)$. For the latter, the empirical equation with three parameters (A, B and n) was used:

$$\Gamma_0(\sigma) = \Gamma_0(\infty) + \frac{A}{(\sigma - B)^n}$$

The linewidth of the macroscopic diamond was taken from the experiment (~30 μm diamond powder): $\Gamma_0(\infty) = 4.2$ cm$^{-1}$ (this value does not include the slit-function).

The particle size distribution $\rho(\sigma)$ was assumed to be Normal with small contribution of log-Normal to describe the fact that nanodiamonds typically contain a few percent of relatively large particles[30]. During the fitting, the varied parameters included the center and widths for distributions, as well as their proportion. Resulting from the fit size distributions are represented graphically on fig. 4.

## Acknowledgements

The authors acknowledge the support from the Grant MOST104-2113-M-009-002.

We would like to thank Prof. Eiji Osawa, NanoCarbon research Institute, Japan, Chandra Prakash Epperla and Prof. H.C.Chang, Institute of Atomic and Molecular Sciences, Academia Sinica, Taiwan and Olga Levinson and Boris Zousman, Ray Techniques Ltd., Israel for the nanodiamond samples.